\documentclass[aps,prb,twocolumn,showpacs,superscriptaddress,groupedaddress,floatfix]{revtex4-1}  
\usepackage{graphicx} 
\usepackage{dcolumn}   
\usepackage{bm}        
\usepackage{amssymb}  
\everymath{\displaystyle}

\begin{document}

\widetext
\leftline{Version 1.0 as of \today}
\leftline{Primary authors: M.~Zybert}
%\leftline{To be submitted to (PRB)}
\leftline{Comment to {\tt zybert.marcin@gmail.com} }
%\centerline{\em D\O\ INTERNAL DOCUMENT -- NOT FOR PUBLIC DISTRIBUTION}

\preprint{APS/123-QED}

\title{Landau levels and shallow donor states in GaAs/AlGaAs multiple quantum wells at mega-gauss magnetic fields}

\author{M. Zybert}
 \affiliation{Centre for Microelectronics and Nanotechnology, University of Rzesz\'ow, Pigonia 1, 35-959 Rzesz\'ow, Poland.}
\author{M. Marchewka}
 \affiliation{Centre for Microelectronics and Nanotechnology, University of Rzesz\'ow, Pigonia 1, 35-959 Rzesz\'ow, Poland.}
\author{E. M. Sheregii}
 \affiliation{Centre for Microelectronics and Nanotechnology, University of Rzesz\'ow, Pigonia 1, 35-959 Rzesz\'ow, Poland.}

\author{D. G. ~Rickel} \affiliation{National High Magnetic Field Laboratory, Los Alamos, New Mexico 87545, USA}
\author{J. B. ~Betts} \affiliation{National High Magnetic Field Laboratory, Los Alamos, New Mexico 87545, USA}
\author{F. F. ~Balakirev} \affiliation{National High Magnetic Field Laboratory, Los Alamos, New Mexico 87545, USA}
\author{M. ~Gordon} \affiliation{National High Magnetic Field Laboratory, Los Alamos, New Mexico 87545, USA}
\author{A. V. Stier} \affiliation{National High Magnetic Field Laboratory, Los Alamos, New Mexico 87545, USA}
\author{C. H. ~Mielke} \affiliation{National High Magnetic Field Laboratory, Los Alamos, New Mexico 87545, USA}

\author{P. Pfeffer} \affiliation{Institute of Physics, Polish Academy of Sciences, Al.
Lotnik\'ow 02-668, Warsaw, Poland}
\author{W. Zawadzki} \affiliation{Institute of Physics, Polish Academy of Sciences, Al.
Lotnik\'ow 02-668, Warsaw, Poland}

\date{\today}

\begin{abstract}
Landau levels and shallow donor states in multiple GaAs/AlGaAs quantum wells (MQWs) are investigated by means of the cyclotron resonance at mega-gauss magnetic fields.  Measurements of magneto-optical transitions were performed in pulsed fields up to 140 T and temperatures from 6 to 300 K. The $14\times14$ \textbf{P}$\cdot$\textbf{p} band model for GaAs is used to interpret free-electron transitions in a magnetic field. Temperature behavior of the observed resonant structure indicates, in addition to the free-electron Landau states, contributions of magneto-donor states in the GaAs wells and possibly in the AlGaAs barriers. The magneto-donor energies are calculated using a variational procedure suitable for high magnetic fields and accounting for conduction band nonparabolicity in GaAs. It is shown that the above states, including their spin splitting, allow one to interpret the observed magneto-optical transitions in MQWs in the middle infrared region. Our experimental and theoretical results at very high magnetic fields are consistent with the picture used previously for GaAs/AlGaAs MQWs at lower magnetic fields.
\end{abstract}

\maketitle

\section{Introduction}

Magneto-donor states in semiconductors have been the subject of sustained experimental and theoretical interest due to their interesting physical properties as well as  important use in the infrared technology \cite{1,2,3,4,5,6,7,8,9,10,11,12}. Magneto-optical and magneto-transport investigations proved to be useful in determining positions of donors in MQWs, which is important for device applications.
It is of interest to verify previous theoretical assumptions derived from experimental data obtain at small magnetic fields and properties of donor centers in MQWs at ultrahigh magnetic fields. One of the important questions is the magnetic field dependence of the optical transition energies for the extreme field range  \cite{13}. This problem is connected with MQWs where the donor centers can exist both in the wells and barriers. In addition to the useful applications of MQWs to infrared photo-detectors  \cite{14,15,16,17}, light emitters  \cite{18,19,20} and cascade lasers \cite{21,22,23,24}, they provide a test system for the study of  electron correlation when the electrons are spatially confined by the potentials of closely spaced multilayers \cite{25,26,27,28,29,30}.

In this paper we present results on the cyclotron resonance (CR) in GaAs/AlGaAs MQWs containing residual Si donors both in the GaAs wells and AlGaAs barriers. The experimental data are obtained in wide range of temperatures. They are described using a sophisticated $14\times14$ band model for free-electron Landau levels and a variational calculations for magneto-donors. The theory takes into account the non-parabolicity and non-sphericity of the conduction band in GaAs which strongly influences all energies at the employed very strong magnetic fields. A connection with the results of other authors is made, both those of Brosak et al \cite{31} who used much lower constant fields, as well as those of Najda et al \cite{10,11} who worked with very high pulsed fields similar to ours.

\section{Experimental results}

Magneto-optical measurements in the infra-red region and pulsed mega-gauss magnetic fields were performed at the National High Magnetic Field Laboratory, Pulsed Field Facility in Los Alamos. Magnetic fields up to 150 T were generated in a single-turn coil discharging a capacity of 250 kJ and inductance of 17.5 nH during 6 $\mu$s. The CR of electrons in the MQW, was excited with the $CO_{2}$ laser radiation at two different wavelengths: $\lambda_{1}=10.59$ $\mu$m ($h\nu=116.7$ meV) and $\lambda_{2}=9.69$ $\mu$m ($h\nu$=128.0 meV) with the power of about 80 mW for each wavelength. The magnetic field B was parallel to the [001] crystal direction of GaAs and maintained perpendicular to the two-dimensional electron gas plane with a special sample holder ensuring the Faraday geometry. 

\begin{figure}[hbp]
	\centering
		\includegraphics[width=0.45\textwidth]{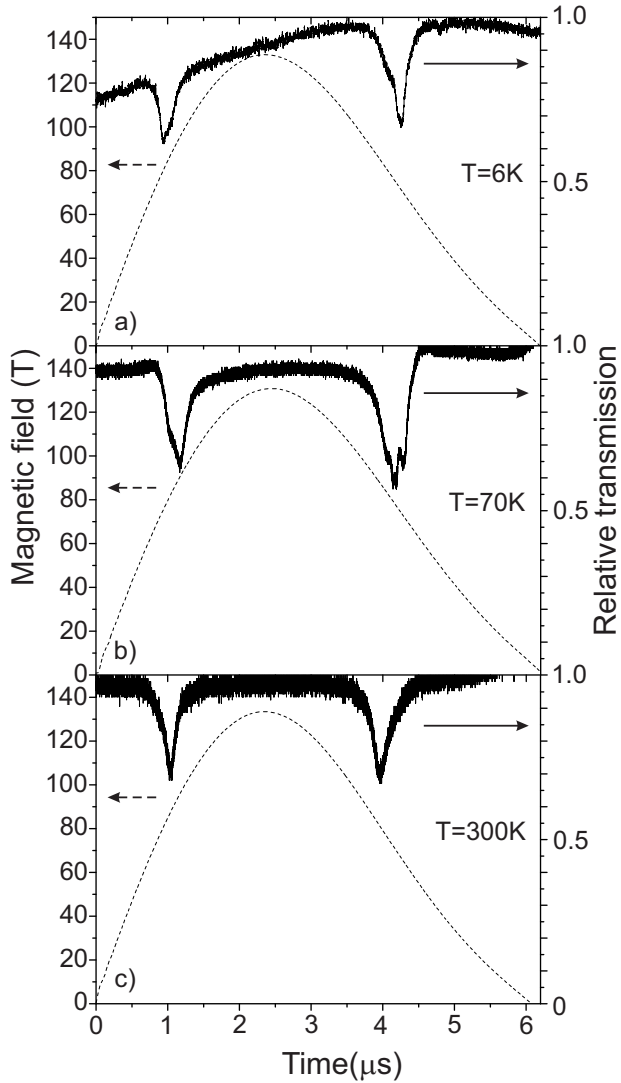}
	\caption{Optical magneto-transmission (solid curves) and magnetic field intensity (dotted curves)versus time for three temperatures and laser wavelength $\lambda_{2}=9.69$ $\mu$m }
	\label{fig:Fig1}
\end{figure}

A HgCdTe detector was used to detect the radiation transmitted through the sample placed within the single-turn coil. The magnetic field induction B was measured at the sample using a $dB/dt$ measuring coil, with an estimated uncertainty not exceeding $\pm$ 3 \%.

\begin{figure}[htp]
	\centering
		\includegraphics[width=0.4\textwidth]{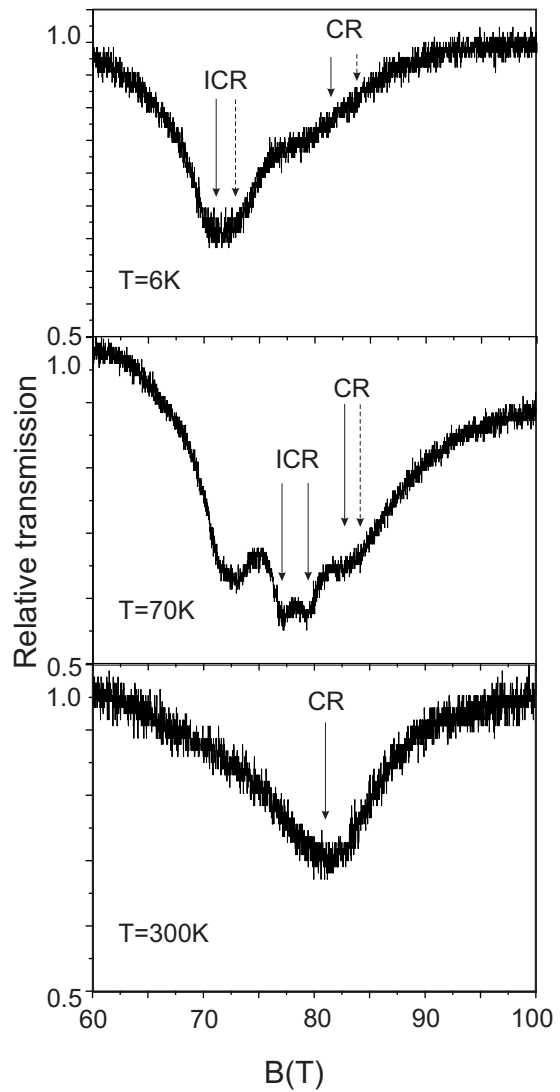}
	\caption{Resonance curves versus magnetic field for $\lambda_{2}=9.69$ $\mu$m at three temperatures obtained at decreased magnetic field. Arrows indicate position ICR and CR peaks,dashed arrows indicate weakly pronounced peaks.}
	\label{fig:Fig2}
\end{figure}

An MQW structure (\#151) prepared specially for our experiment consisted of ten GaAs QWs and eleven $Al_{x}Ga_{1-x}As$ barriers grown on GaAs substrate \cite{32}. The well thickness was 18 atomic layers (AL, about 10 nm) while the width of barriers was 9 AL (about 5 nm). Magneto-transport measurements\cite{30} at temperatures from 1.6 K to 4.2 K determined the electron density of $5\cdot10^{11} $ cm$^{−-2}$ and electron mobility of about  $5\cdot10^{4} $ cm$^{2}/$Vs .

\begin{figure}[htp]
	\centering
		\includegraphics[width=0.4\textwidth]{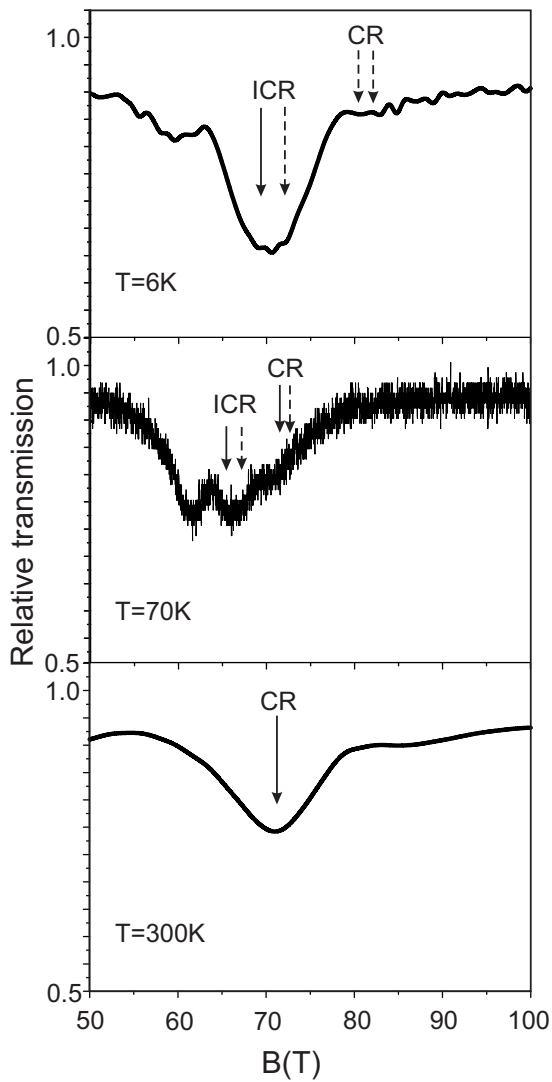}
	\caption{Resonance curves versus pulsed magnetic field recorded for $\lambda_{2}=10.59$ $\mu$m at three temperatures obtained at decreased magnetic field. Arrows indicate positions of ICR and CR peaks, dashed arrows indicate weakly pronounced peaks.}
	\label{fig:Fig3}
\end{figure}

Magneto-transmission curves versus time recorded for $\lambda_{1}=9.69$ $\mu$m at three temperatures (6 K, 70 K and 300 K) are presented in Fig. 1. Magnetic field curves versus time are given on the same figures by dotted curves. Each transmission spectrum has clearly visible resonance minima. The resonance peaks were reproducible while measured for increasing and decreasing magnetic fields. However, shifts to higher magnetic fields were found on curves corresponding to increasing fields. This is illustrated in Table 1. This effect could be explained by the electron relaxation processes associated which short pulse duration \cite{33}.
The values of resonance fields for increasing and decreasing field runs were averaged 

\begin{equation}
B^{exp}_{r}=\frac{1}{2}\left(B^{incr}_{r}-B^{decr}_{r}\right)
\end{equation}

Miura and coworkers \cite{10,11} showed in his experiments on CR at very high magnetic fields how temperature affects the structure of resonant peaks. At low temperatures the resonances are dominated by impurity transitions (ICR) related to magneto-donors, as the temperature increases the free electron transitions begin to dominate. When the source has a fixed radiation frequency, the donor related transitions occur at lower magnetic fields. This behavior is illustrated in our Figs. 1-3. At T = 6 K the free-electron resonance is negligible, while the measurements at 70 K show both CR and ICR. Finally, at T = 300 K one sees only the free-electron CR. 
At about 76.5 T ($\pm$ 3 T) a stronger peak is visible at 6 K and $\lambda_{1}=9.69$ $\mu$m, whereas, a weaker one is at about 85.5 T ($\pm$ 4 T). The peaks are split into two, more visible splitting is observed on curves for decreasing fields, see Fig. 1. 

\begin{table*} [htp]

\caption{\label{tab:table1}. Positions of resonance fields for increasing and decreasing field runs, as well as  average values. The asterisks mark weaker peaks}

\begin{tabular}{|c|c|c|c|c|}
\hline
T(K)  & E(meV)&   Incr. peak position	(B)	& Decr. peak position  (B)	& Aver. peak position (B) \\ \hline 
6			& 116.7	& 66												& 60												& 63 \\ 
6			& 116.7	& 71												& 68												& 69.5 \\ 
6			& 116.7	& 73												& 70												& 71.5* \\ 
6			& 116.7	& 82												& 80												& 81* \\ 
6			& 116.7	& 83												& 81												& 82* \\ \hline 
6			& 128.0	& 75												& 65												& 70 \\ 
6			& 128.0	& 81.5											& 71.5											& 76.5 \\ 
6			& 128.0	& 84												& 73												& 78.5* \\ 
6			& 128.0	& 89												& 82												& 85.5 \\ 
6			& 128.0	& 91												& 84												& 87.5* \\ \hline 
70		& 116.7	& 62												& 58												& 60 \\ 
70		& 116.7	& 72												& 65												& 68.5 \\ 
70		& 116.7	& 75												& 61												& 71 \\ 
70		& 116.7	& 79												& 71												& 75 \\ 
70		& 116.7	& 80.5											& 72.5											& 76.5* \\ \hline
70		& 128.0	& 77												& 70												& 73.5* \\ 
70		& 128.0	& 79												& 72												& 75.5 \\ 
70		& 128.0	& 86.5											& 77.5											& 81.5 \\ 
70		& 128.0	& 87.5											& 79.5											& 83.5 \\ 
70		& 128.0	& 91												& 83												& 87.5 \\ 
70		& 128.0	& 92.5											& 84.5											& 88.5* \\ \hline 
300		& 116.7	& 74												& 71												& 72.5 \\ \hline 
300		& 128.0	& 87												& 85												& 86 \\ \hline 

\end{tabular}

\end{table*}

Increasing the temperature to 70 K causes appearance of more complex resonant structure: three strong resonance peaks are observed at (87 $\pm$ 4) T, (81.5 $\pm$ 4) T and (75.5 $\pm$ 3) T. The central resonance peak at 70 K is more clearly split, as seen in Fig. 2, but other peaks are split as well. In contrast, the curve at 300 K shows a strong peak at the field (86 $\pm$ 4)T. The curves for $\lambda_{1}=10.59$ $\mu$m at different temperatures are presented in Fig. 3. One observes the same resonances as those shown in Figs. 1 and 2, but somewhat shifted toward smaller fields. The experimental values and their averages for increasing and decreasing fields are presented for both wavelengths in Table 1.

\section{Theory}

It was shown by magneto-optical studies of the conduction band of GaAs both at low and high magnetic fields that, in addition to the free-electron Landau states, one usually deals in this material with residual Si donors. This is also the case in our studies, as indicated by preliminary inspection of the resonant peaks. They cannot be explained by the free electron cyclotron resonance alone, also when one takes into account the fact that both $n=0$ and $n=1$ Landau levels (LLs) are spin split and, due to band nonparabolicity, the spin splitting of each LL is different. Thus, we have to consider both free-electron and magneto-donor (MD) optical transitions. The free-electron LLs can be described quite precisely including band nonparabolicity and nonsphericity, the description of MD energies is more complicated and we must recourse to approximate procedures.

\subsection{Free electrons}

The description of free-electron LLs in the nonparabolic and nonspherical conduction band of GaAs was worked out and verified in detail by Pfeffer and Zawadzki \cite{34}, so we will give here only a short summary of this work and use its results. GaAs is a medium-gap material, so that in order to describe correctly its conduction band it is not enough to apply the standard three-level \textbf{P}$\cdot$\textbf{p} model used for narrow-gap semiconductors. Thus, a five-level \textbf{P}$\cdot$\textbf{p} model is used which includes in addition two higher conduction levels.One takes at the $\Gamma$ point of the Bruillouin zone two $\Gamma^{v}_{15}$ valence levels,$\Gamma^{c}_{1}$  conduction  level, and two $\Gamma^{c}_{15}$  conduction levels. This gives, including degeneracies and spins, 14 states. The initial multi-band \textbf{P}$\cdot$\textbf{p} set for carriers in the presence of an external magnetic field is:

\begin{equation}
\resizebox{.9\hsize}{!}{${\sum_{l}\left[\left(\frac{\textbf{P}^{2}}{2m_{0}}+E_{l0}-E\right)\delta_{l'l}+\frac{\textbf{p}_{l^{'}l}\cdot \textbf{P}}{2m_{0}}+\mu_{B}B\cdot \sigma_{l'l}+H_{l'l}^{S.O.}\right]f_{l}}=0$}
\end{equation}

where $\textbf{P}=\textbf{p}+e\textbf{A}$ is the kinetic momentum, \textbf{A} is the vector potential of magnetic field \textbf{B}, $E_{l0}$ are the band-edge energies, $\textbf{p}_{l'l}$ are the interband matrix elements of momentum, $\sigma_{l'l}$ those of the spin operators and $H_{l'l}^{S.O.}$ those of the spin-orbit interaction.The summation runs over 14 bands. Equation (2) represents a set of coupled differential equations for the envelope functions $f_{l}$. Far-band contributions are included using the perturbation theory up to the $P^{2}$ terms. If the considered energy bands were spherical, one could find solutions of the set (2) by a column of single harmonic oscillator functions. However, an interaction of the two higher conduction levels with the two lowest valence levels of the set results in a slight nonsphericity of the bands, including the $\Gamma^{c}_{6}$ conduction band of our interest. To account for this feature, one looks for solutions of the problem (2) in the form of sums of harmonic oscillator functions \cite{35}. It turns out that LLs have somewhat different energies for [001], [110], [111] field orientations. This is of particular importance for the spin splittings which can change signs from negative to positive as the magnetic field increases. 
When computing energies we have to use the material parameters for the five-level model. We take the conduction band-edge values of  $m^{*}_{0} = 0.066 m_{0}$ and $g^{*}_{0} = - 0.44$, as determined by the cyclotron and spin resonances. The mass value includes the so called polaron contribution, i.e. the effect of non resonant electron-polar phonon interaction. This interaction increases the effective mass at low magnetic fields according to the relation

\begin{equation}
m^{*}(exp)=m^{*}_{0}\left(1 - \frac{\alpha}{6}\right)^{-1}
\end{equation}

where $\alpha$ is the polar coupling constant. Knowing the value of $\alpha = 0.085$ and the experimentally measured mass $m^{*}(exp)$, one determines the “bare” mass  $m^{*}_{0} = 0.0651 m_{0}$ which should be used in the Landau level calculations for very high magnetic fields at which the optic phonons do not contribute. We use the following values of experimental gaps: $E_{0}=-1.519 $eV and the matrix elements of momentum: $E_{P_{0}}=27.86 $eV, in the standard units $E_{P}=2m_{0} P^{2}/\hbar^{2}$. The Luttinger valence-band parameters
resulting from the interaction of far bands with the $\Gamma^{v}_{15}$ bands are: $\gamma^{L}_{1}=7.80$, $\gamma^{L}_{2}=2.46$, $\gamma^{L}_{3}=3.30$, $\kappa^{L}=2.03$ \cite{34}.

The basic matrix that has to be computed for a given LL \textit{n} and specific spin orientation has  dimensions $7\times7$. However, the basic $7\times7$ matrices for different \textit{n} are coupled by the nonspherical terms into matrices of higher dimensions. In order to obtain the sufficient precision for the field \textbf{B} $\parallel$ [001] we truncate the matrices at the dimension $35\times35$ and compute their eigen energies. The calculated energies exhibit nonlinear dependence of LLs on B due to band nonparabolicity. The most striking feature is the change of spin splitting from negative at low fields (expressed by the negative values of spin\textit{ g}-factors) to positive at high fields. The change of sign occurs at lower B intensities for LLs with higher \textit{n}, see Ref. \cite{34}. This is illustrated schematically in Fig.4 which shows that for $n=0$ the \textit{g}-value is negative ($0+$ state is lower than $0-$), while for $n=1$ the g-value is positive ($1-$ state is lower than $1+$). In the calculations we do not change the energy gap of GaAs for temperatures between 6 K and 70 K because, as follows from Ref. \cite{36}, the change of energy gap due to dilatation in this material is negligible in the low temperature range.

\subsection{Magneto-donors}

In order to treat magneto-donor (MD) energies at the comparable level of precision, one would need to write down the donor potential in the 14 diagonal terms of the initial matrix (2) and deal with the resulting eigenvalue problem. This is not a tractable task, so we have to recourse to approximate solutions. A key parameter in the MD problem is ratio of the binding donor energy to the magnetic energy, i.e:

\begin{equation}
\gamma=\frac{h\omega_{c}}{2Ry^{*}}
\end{equation}

where $\omega_{c}=eB/m^{*}_{0}$ and $Ry^{*}=m^{*}e^{4}/2\kappa^{2} \hbar^2$ is the effective Rydberg. In GaAs there is $Ry^{*}=5.9 $meV, so that at a magnetic field of B 86 T we have $\gamma=13.6$. It follows from the work of Brozak et al \cite{31} that at values $\gamma>6$ one can treat the MD problem in a quantum well with the magnetic field parallel to the interfaces, i.e. in the Faraday configuration, as a problem in the bulk. This considerably facilitates our task. 

\begin{figure}[htp]
	\centering
		\includegraphics[width=0.45\textwidth]{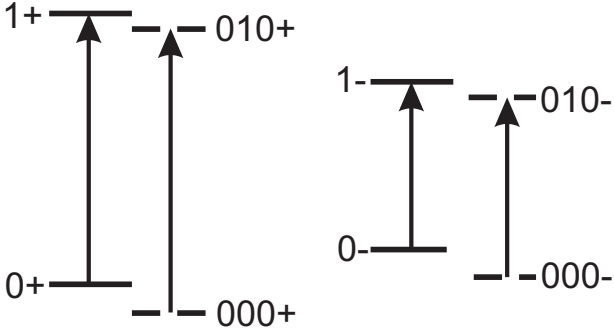}
	\caption{Cyclotron resonance and MD cyclotron resonance transitions for both spin orientation}
	\label{fig:Fig4}
\end{figure}

We want to solve the MD problem using variational procedure. Since the formalism of $14\times14$ matrix is not tractable for this purpose,we imitate the nonparabolicity of the conduction band by employing a two-band model with an \textit{effective energy gap} $\epsilon^{*}_{g}$.Thus, we take the effective gap value which gives the same nonparabolicity as the $14\times14$ band procedure. It was shown in ref. \cite{11} that the value of such a gap for GaAs is 0.98 eV. Then the two-band equation (omitting spin) is

\begin{equation}
E= \frac{-\epsilon^{*}_{g}}{2}+\left[\left(\frac{\epsilon^{*}_{g}}{2}\right)^{2}+\epsilon^{*}_{g}\left\langle K\right\rangle \right]^{1/2}+\left\langle U\right\rangle
\end{equation}

where the variational averages of kinetic and potential parts of the MD energy are, correspondingly (in the cylindrical coordinate system)

\begin{equation}
\left\langle K\right\rangle = \left\langle \Psi_{NM\beta}\left|-\nabla^{2} - i\gamma \frac{\partial}{\partial\varphi} + \frac{\gamma^{2}\rho^{2}}{4}\right|\Psi_{NM\beta}\right\rangle,
\end{equation}

\begin{equation}
\left\langle U\right\rangle = \left\langle \Psi_{NM\beta}\left|\frac{-2}{\left(\sqrt{z^{2}+\rho^{2}}\right)}\right|\Psi_{NM\beta}\right\rangle.
\end{equation}

The energies are in effective Rydbergs and lengths in the effective Bohr radii. The potential energy in Eq. (5) stands outside the square root since, in the multi-band \textbf{P}$\cdot$\textbf{p} matrix, the potential always appears in diagonal terms together with the energy. One calculates the variational averages of $\left\langle U\right\rangle$ and $\left\langle K\right\rangle$  and than minimizes the energy of Eq. (5). However, we cannot hope to get sufficiently precise absolute MD energies from the above variational and simplified band structure procedures to be compared with the precise free electron energies. For this reason we calculate from Eq. (5) only shifts of the MD energies,as counted from the free-electron energies. The calculation of the shifts amounts to separate evaluation of the variational energies according to Eq.(5) and their comparison with the free-electron energies according to the same two-band model, by putting in Eq. (5) $\left\langle U\right\rangle$  = 0 and $\left\langle K\right\rangle =2\gamma(n+1/2)$, i.e. the energy of free-electron LL \textit{n}.

As to spin contributions to the energies, the two-band equation of the type (5) can not reproduce the change of signs of the spin splittings mentioned above. In this situation, we assume that the spin splitting of MD energies is the same as that of the free-electron energies calculated from the $14\times14$ scheme. Thus, in order to obtain the complete MD energies, we shift the calculated free-electron Landau levels (which include the spin) by the above mentioned amounts not depending on the spin.The assumption of identical spin splittings for LLs and MD energies is well justified since the energy differences between free and bound electron states are much smaller than their absolute energies at high fields.
Since we deal with very high magnetic fields, expressed by the high values of $\gamma$, we can use in the variational calculations of MD energies one-parameter trial functions proposed by Wallis and Bowlden \cite{35}. These functions express the fact that, in the MD state, component of the motion transverse to magnetic field is almost equal to that of the magnetic radius for a free electron, so that one varies only the longitudinal component, see \cite{11}. Here $\lambda$ is variational parameter, $L_{N}^{M}$ are associated Laguerre polynomials and $P_{β}(z)$ are orthogonal polynomials. The quantum numbers are: $N = 0,1,2, …, M = …-1, 0, 1,…, β = 0, 1, 2,… $We need only $P_{0}(z)=\left(\gamma\lambda/2\pi\right)^{\left(1/4\right)}$. 
The Landau level number to which a given MD state "belongs" is $n=N+1/2(M+|M|)$.
For the MD states of our interest we take explicitly

\begin{equation}
\Psi_{000}=C\cdot e^{-\frac{\rho}{2}+1} \rho L^{0}_{0}(\rho)  \left(\frac{\gamma\lambda}{2\pi}\right)^{\frac{1}{4} } e^{-\frac{1}{4} \gamma \lambda z^{2}}
\end{equation}

\begin{equation}
\Psi_{010}=C\cdot e^{i\varphi} e^{-\frac{\rho}{2}+1} \rho^{-\frac{1}{2}} L^{1}_{0}(\rho) \left(\frac{\gamma\lambda}{2\pi}\right)^{\frac{1}{4}} e^{-\frac{1}{4} \gamma \lambda z^{2}} 
\end{equation}

where $L^{0}_{0}(\rho)=L^{1}_{0}(\rho)=1$. The normalization coefficients \textit{C} and the variational parameters $\lambda$ are different for each function.
Thus the complete theory includes in the first step precise calculation of the free-electron energies with the use of $14\times14$ \textbf{P}$\cdot$\textbf{p} formalism. These are directly used to interpret the free electron data. Next, we calculate variational MD energies from the two-band equation (5) with the effective energy gap and calculate the MD energy shifts using the same two-band equation (5) for free electrons with the same effective gap. Finally, the obtained shifts are subtracted from the exactly calculated free-electron energies and used to interpret the experimental results for magneto-donors.

\section{Comparison of experiment with theory}

In Fig.5 we show all observed resonance points and lines calculated according to the presented theory for magnetooptical transitions between the free-electron LLs $n = 0$ and $n = 1$, as well as MD states (000) and (010). The spin splittings are calculated for free electrons, as explained above. Experimental positions of the observed resonances at 6 K are indicated by black squares. It is seen that the observed central stronger peak should be attributed to the  transition between the MD states.The uncertainties of our experimental points are determined by different resonance positions for increasing and decreasing field runs, as indicated in Table 1. In general, the agreement between our experiment and theory is quite good for both LL and MD spin doublets shown in Fig.5. The spin \textit{g}-value for conduction electrons in GaAs is known to be very small, but at our very high magnetic field it results in sizable splittings. The data show that the spin splittings of LL and MD transitions are quite similar indicating that our assumption in this respect was reasonable. Finally, a good agreement of experiment and theory for free-electron transitions confirms indirectly but convincingly that, indeed, the polaron corrections to the effective electron mass are absent at high magnetic fields. 

\begin{figure}[htp]
	\centering
		\includegraphics[width=0.45\textwidth]{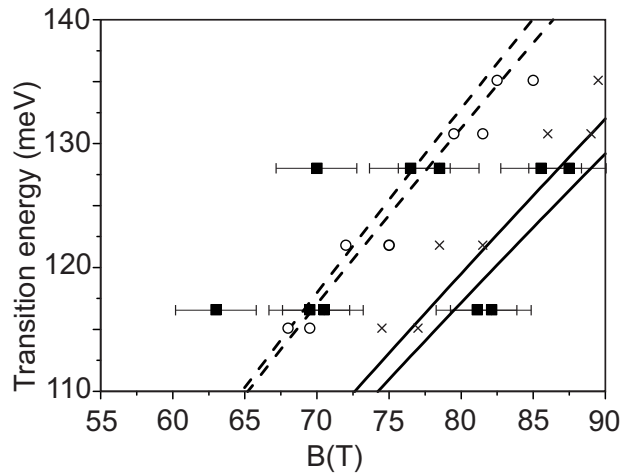}
	\caption{Magnetooptical transition energies calculated including spin splitting according theory presented in chapter III and experimental data obtained at 6 K: full squares
are our data, circles and crosses are data of Ref. 11}
	\label{fig:Fig5}
\end{figure}

There appear two additional resonances on the lower field side whose origin is not clear. Huant et al \cite{3,4,5,38} observed in GaAs/AlGaAs quantum wells  magneto-optical transitions ascribed to donors in AlGaAs barriers \cite{3,4,5,38,39}. According to calculations \cite{38} and observations reported in \cite{39} the energies of transitions for MD in barriers are slightly higher than the cyclotron resonance. This would correspond in our experiments with fixed laser frequencies to resonances on weaker field sides, in agreement with our observations. On the other hand, the MD donors in the center of a barrier  have distinctly smaller transition energies than those in the center of a well. Because donors in our MQW are in wells and barriers we presume that these resonances correspond to optical transitions of electrons in barriers.

\begin{figure}[htp]
	\centering
		\includegraphics[width=0.45\textwidth]{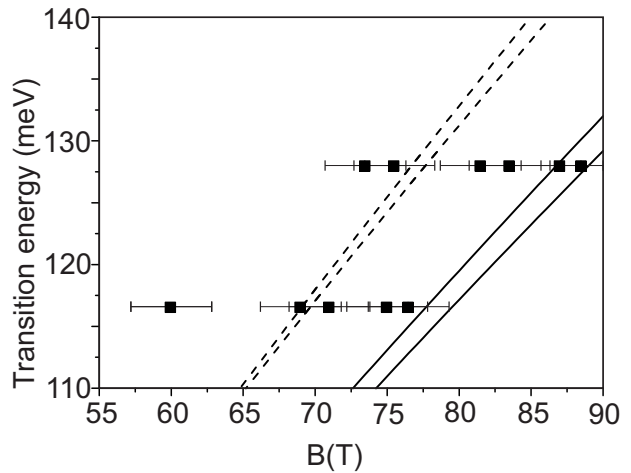}
	\caption{Positions of resonance peaks at 70 K for wavelengths $\lambda_{1}$ and $\lambda_{2}$ and theoretical lines versus magnetic field.}
	\label{fig:Fig6}
\end{figure}

Weak minimum recorded at higher magnetic fields, as it is seen from Figs. 1-3 is caused by CR in QW although agreement with the theoretical position of this resonance is worse. Temperature dynamic of observed peaks confirms this interpretation – while, at higher temperature (70 K, see Figs. 1-3) this peak is stronger and at 300 K it becomes a single peak. 
Resonance recorded from the smaller fields at 70 K (see Fig. 4) should be referred to the MD transitions in barriers. The reduced non-ionized donors in barriers cause this transitions as it have been shown in Refs. \cite{3,4,5,39} in the region of smaller magnetic fields. In the case of the MD transitions in the AlGaAs barriers the spin splitting is smaller and is outside experimental resolution of 1.5 T.
In Fig. 6 we show the data taken at T = 70 K and unchanged theoretical lines. There appears a doublet for the higher frequency between the theoretical lines whose origin is unclear. In principle higher temperatures activate free-electron transitions but there are no free-electron states to produce this doublet. There is one resonance at lower magnetic fields, similar to those indicated in Fig. 5. 
Nevertheless, we find that the overall agreement between the experiment and theory is quite reasonable, confirming our simplified treatment of magneto-donor energies. There appear few additional unexplained resonances which can be attributed to inhomogeneous character of multiple GaAs/AlGaAs quantum wells doped with Si.

\section{Summary}

Very high pulsed magnetic fields up to 140 T were used to study cyclotron resonance and magneto-donor optical transitions in multiple GaAs/AlGaAs quantum wells in the Faraday configuration. The magneto-optical spectra taken  at T = 6, 70 and 300 K exhibit different details as the temperature changes. The observed free-electron magneto-optical transitions were described by the \textbf{P}$\cdot$\textbf{p} theory including 14 energy bands that accounts for the nonparabolicity and nonsphericity of the conduction band in GaAs. Shifts of magneto-donor energies with respect to the free-electron energies were calculated by a variational procedure taking into account effective nonparabolicity of the conduction band in GaAs. The applied theory successfully describes our data, as well as data of other authors quoted for comparison. A possible origin of a few unexplained resonances is discussed. Our study confirms the general picture of both free electron and magnetodonor states in GaAs/AlGaAs multiple quantum wells in very wide range of magnetic fields which is important for various applications.

\section{ACKNOWLEDGEMENT}

This work is supported by National Science Foundation -- Cooperative Agreement No. DMR-1157490, the State of Florida, and the U.S. Department of Energy. We are grateful to NFS and LANL for opportunity to preform MGFCR experiment.


\begin{thebibliography}{99}
\bibitem {1} D. M.Larsen, Phys. Rev. Lett. \textbf{42}, 742 (1979)
\bibitem {2} D. M.Larsen, Phys. Rev. B \textbf{20}, 5217 (1979)
\bibitem {3} S.Huant, S. P. Najda, and B.Etienne, Phys. Rev. Lett. \textbf{65}, 1486 (1990)
\bibitem {4} S. Huant, A. Mandray and B. Etienne, Physica Scripta \textbf{45}, 145 (1992)
\bibitem {5} S. Huant, A. Manday, J. Zhu, S. G. Louie, T. Pang, B. Etienne, Phys. Rev. B 48, 2370 (1993)
\bibitem {6} J.-P. Cheng, Y. J. Wang, B.D. McCombe, W. Schaff, Phys. Rev. Lett. \textbf{70}, 489 (1993)
\bibitem {7} E.R. Mueller, D.M. Larsen, J. Walsman, W. D. Goodhue, Phys. Rev. Lett. \textbf{68}, 2204 (1992)
\bibitem {8} W. Zawadzki, J. Wlasak, in: J. T. Devreese (ed.) \textit{Theoretical Aspects and New Developments in Magneto-Optics}, Springer, N-Y,	1980, pp. 347-390
\bibitem {9} W. Zawadzki, in: Landau Level Sectroscopy, Ed. G. Landwehr, E. I. Rashba, North-Holland, pp. 679-776 (1991)
\bibitem {10} S. P. Najda, S. Takeyama, N. Miura, P. Pfeffer , Zawadzki, Phys. Rev. B \textbf{40}, 6189 (1989)
\bibitem {11} W. Zawadzki, P. Pfeffer, S. P. Najda, H. Yokoi, S. Takeyama, and N. Miura, Phys. Rev. B \textbf{49}, 1705 (1994)
\bibitem {12} A. B. Dzyubenko, A. Manday, S. Huant, A.Yu. Sivachenko, B. Etienne, Phys. Rev. B \textbf{50},  4687 (1994)
\bibitem {13} L. G. Booshehri, C. H. Mielke, D. G. Rickel, S. A. Crooker, Q. Zhang, L. Ren, E. H. Hroz, A. Rustagi, C. J. Stanton, Z. Jin, Z. Sun, Z. Yan, J. M. Tour, and J. Kono, Phys. Rev. B \textbf{85}, 205407 (2012)
\bibitem {14} M. O. Manasreh, Gary L. McCoy, Semiconductor Quantum Wells and Superlattices for Long-Wavelength Infrared Detectors, ARTECH, Boston (1993)
\bibitem {15} S. M. Ramey , R. Khoie , IEEE Transactions on Electron Devices \textbf{50}, 1179(2003). 
\bibitem {16} V. Ryzhii, M. Ryzhii, Phys. Rev. B \textbf{62}, 10292 (2000).
\bibitem {17} L. Höglund, K. F. Karlsson, P. O. Holtz, H. Pettersson, M. E. Pistol, Q. Wang, S. Almqvist, C. Asplund, H. Malm, E. Petrini, and J. Y. Andersson,Phys. Rev.B \textbf{82}, 035314 (2010)
\bibitem {18} C. Chen, N. Braidy, C. Couteau, C. Fradin, G. Weihs, R. LaPierre, NanoLetters \textbf{8} (2),495–499 (2008)
\bibitem {19} G. Franssen, P. Perlin, and T. Suski, Phys. Rev. B \textbf{69}, 045310 (2004)
\bibitem {20} L. Rigutti, A. Castaldini, and A. Cavallini, Phys. Rev. B \textbf{77}, 045312 (2008)
\bibitem {21} M. Za\l u\.zny, Phys. Rev. B \textbf{43}, 4511 (1991).
\bibitem {22} F. Szmulowicz, M. O. Manasreh, C. E. Stutz, and T. Vaughan, Phys. Rev. B \textbf{50}, 11 618 (1994).
\bibitem {23} J. Faist, F. Capasso, D. L. Sivco, C. Sirtori, A. L. Hutchinson and A. Y. Cho, Science,  \textbf{264}, 553 (1994) 
\bibitem {24} P. Rauter, T. Fromherz, N. Q. Vinh, B. N. Murdin, G. Mussler, D. Grutzmacher, and G. Bauer, Phys. Rev. Letters \textbf{102}, 147401 (2009)
\bibitem {25} Danhong Huang, M.O. Monosrech, Phys. Rev. B \textbf{54}, 2044 (1996)
\bibitem {26} C. Negrevergne, T. S. Mahesh, C. A. Ryan, M. Ditty, F. Cyr-Racine, W. Power, N. Boulant, T. Havel, D. G. Cory, and R. Laflamme, Phys. Rev. Letters \textbf{96}, 170501 (2006)
\bibitem {27} D. P\l och, E.M. Sheregii, M. Marchewka, M. Woźny and G. Tomaka, Phys. Rev. B \textbf{79}, 195434 (2009)
\bibitem {28} M. Marchewka, E.M. Sheregii, I. Tralle, A. Marcelli, M. Piccinini and J. Cebulski, Phys. Rev. B \textbf{80}, 125316 (2009)
\bibitem {29} M. Marchewka, E.M. Sheregii, I. Tralle, D. P\l och, G. Tomaka, M. Furdak, A. Kolek, A. Stadler, K. Mleczko, D. \.Zak, W. Strupi\'nski, A. Jasik, R. Jakie\l a,  Physica E \textbf{40}, 894 (2008)
\bibitem {30} M. Zybert, M. Marchewka, G. Tomaka and E.M. Sheregii, Physica E \textbf{44}, 2056 (2012)
\bibitem {31} B. Brozak, B.D. McCombe, D. M. Larsen, Phys. Rev. B \textbf{40} 1265 (1989)

\bibitem {32} A. Jasik, A. Wnuk, J. Gaca, M. W\'ojcik, A. W\'ojcik-Jedli\'nska, J. Muszalski, W. Strupi\'nski, Journal of CrystalGrowth  \textbf{311}, 4423 (2009) 
\bibitem {33} S. Hansel, C. Puhle, M. von Ortenberg, E. Huseynov, Physica B: Condensed Matter  \textbf{346},479 (2004)
\bibitem {34} P. Pfeffer, W. Zawadzki, Phys. Rev. B \textbf{53}, 12813 (1996) 
\bibitem {35} V. Evtuhov, Phys. Rev. \textbf{125}, 1869 (1962)
\bibitem {36} W. Zawadzki, P. Pfeffer, R. Bratschitsch, Z. Chen, S. T. Cundiff, B. N. Murdin, C. R. Pidgeon, Phys. Rev. B \textbf{78}, 245203 (2008) 
\bibitem {37} R. F. Wallis, H. J. Bowlden  J. Phys. Chem. Solids \textbf{7}, 78 (1958)
\bibitem {38} S. Huant, A. Mandray, B. Etienne, Solid State Commun. \textbf{93}, 435 (1995)
\bibitem {39} E. Glaser, B. V. Shanabrook, R. L. Hawkins, W. Beard, J. M. Mercy, B. D. McCombe, D. Musser Phys. Rev. B \textbf{36}, 8185 (1987)


\end{thebibliography}
\end{document}